\documentclass[twocolumn]{aa} %
\pdfoutput=1
\usepackage{graphicx}
\usepackage{txfonts}
\usepackage{txfonts}
\usepackage{amsmath}
\usepackage{xcolor}
\usepackage{ulem}
\usepackage[utf8]{inputenc}
\usepackage{multirow}

\usepackage[]{appendix}

\newcommand{\snks}[0]{(\mathrm{S/N})_{\mathrm{KS}}}%

\usepackage{natbib,twoopt}
\usepackage[breaklinks=true]{hyperref} 
\bibpunct{(}{)}{;}{a}{}{,}             
\makeatletter
  \newcommandtwoopt{\citeads}[3][][]{\href{http://adsabs.harvard.edu/abs/#3}%
    {\def\hyper@linkstart##1##2{}%
     \let\hyper@linkend\@empty\citealp[#1][#2]{#3}}}
  \newcommandtwoopt{\citepads}[3][][]{\href{http://adsabs.harvard.edu/abs/#3}%
    {\def\hyper@linkstart##1##2{}%
     \let\hyper@linkend\@empty\citep[#1][#2]{#3}}}
  \newcommandtwoopt{\citetads}[3][][]{\href{http://adsabs.harvard.edu/abs/#3}%
    {\def\hyper@linkstart##1##2{}%
     \let\hyper@linkend\@empty\citet[#1][#2]{#3}}}
  \newcommandtwoopt{\citeyearads}[3][][]%
    {\href{http://adsabs.harvard.edu/abs/#3}
    {\def\hyper@linkstart##1##2{}%
     \let\hyper@linkend\@empty\citeyear[#1][#2]{#3}}}
\makeatother

\begin{document} 


\title{Efficiently combining $\alpha$ CenA multi-epoch \\ high-contrast imaging data \\[1ex] \large Application of K-Stacker to the 80 hrs NEAR campaign}
\titlerunning{Efficiently combining $\alpha$ CenA multi-epoch high-contrast imaging data}

\author{H. Le Coroller \inst{1}
          \and
          M. Nowak \inst{2, 3}
          \and
          K. Wagner \inst{4, 5}
          \and
          M. Kasper \inst{6}
          \and
          G. Chauvin  \inst{7}
          \and 
          C. Desgrange \inst{8, 9}
          \and 
          S. Conseil \inst{1}
        }
        
   \institute{
    Aix Marseille Univ., CNRS, CNES, LAM, Marseille, France.
    \email{herve.lecoroller@lam.fr}
    \and
    Institute of Astronomy, University of Cambridge, Madingley Road, Cambridge CB3 0HA, UK.
    \and
    Kavli Institute for Cosmology, University of Cambridge, Madingley Road, Cambridge CB3 0HA, UK.
    \and
    Dept. of Astronomy and Steward Observatory, University of Arizona, Tucson, AZ, USA.
    \and
    NASA Hubble Postdoctoral Fellowship Program - Sagan Fellow.
    \and
    European Southern Observatory, Garching bei München, Germany.
    \and
    Université Côte d'Azur, Observatoire de la Côte d'Azur, CNRS, Laboratoire Lagrange, Nice, France
    \and 
    Univ. Grenoble-Alpes, CNRS, IPAG, 38000 Grenoble, France.
    \and
    Max-Planck-Institut für Astronomie, Königstuhl 17, 69117 Heidelberg, Germany
    }


 
   \date{}

 
  \abstract
   {Keplerian-Stacker is an algorithm able to combine multiple observations acquired at different epochs taking into account the orbital motion of a potential planet present in the images to boost the ultimate detection limit. In 2019, a total of 100 hours of observation were allocated to VLT/VISIR NEAR, a collaboration between ESO and Breakthrough Initiatives, to search for low mass planets in the habitable zone of the $\alpha$ Cen AB binary system. A weak signal ($\mathrm{S/N} \sim{} 3 $) was reported around $\alpha$ Cen A, at a separation of $\simeq{}1.1\,\mathrm{au}$ which corresponds to the habitable zone. 
  }
   {We aim at determining if K-Stacker also detects the low-mass planet candidate found by NEAR team, with similar orbital parameters. We also aim at searching for additional potential candidate around $\alpha$ Cen A by utilizing the orbital motion to boost the signal, and in general at putting stronger constraints on the presence of other planets in the system.}
   {We have re-analysed the NEAR data using K-Stacker. This algorithm is a brute-force method able to find planets in time series of observations and to constrain their orbital parameters, even if they remain undetected in a single epoch.}
   {We scanned a total of about $3.5\times{}10^5$ independent orbits, among which about 15 \% correspond to fast moving orbits on which planets cannot be detected without taking into account the orbital motion. We find only a single planet candidate, which matches the C1 detection reported in \cite{2021NatCo..12..922W}. However, since this constitutes a re-analysis of the same data set, more observations will be necessary to confirm that C1 is indeed a planet and not a disk or other data artifact. Despite the significant amount of time spent on this target, the orbit of this candidate remains poorly constrained due to these observations being closely distributed in 34 days. We argue that future single-target deep surveys would benefit from a K-Stacker based strategy, where the observations would be split  over a significant part of the expected orbital period to better constrain the orbital parameters. 
   }
    {This application of K-Stacker on high contrast imaging data in the mid-infrared demonstrates the capability of this algorithm to aid in the search for Earth-like planets in the habitable zone of the nearest stars with future instruments of the E-ELT such as METIS.}
   \keywords{methods: observational - methods: data analysis - instrumentation: adaptive optics - instrumentation: high angular resolution - planets and satellites: dynamical evolution and stability - stars: individual: Alpha Cen A}

   \maketitle
   
%

\section{Introduction}

High contrast imaging (HCI) techniques allow to search for planets in a relatively large field of view, to measure their luminosities (intrinsic or reflected), to determine their astrometric positions, and to determine their orbits. In HCI, observations aim at spatially separating the planet from its stellar host, thus allowing to record spectra of the detected planets, either using an Integral Field Unit (IFU), or by optically coupling the imager to a high-resolution spectrograph \citepads{2021A&A...646A}. 

So far, most high contrast imaging instruments (GPI, \citeads{2014PNAS..11112661M} ; SCExAO, \citeads{2015PASP..127..890J}; SPHERE, \citeads{2019A&A...631A.155B}) have been operating in the near-infrared ($\approx 1 - 3 \, \mu$m). This is an optimal choice when looking for young ($\le 100\,\mathrm{Myr}$) giant (a few $M_\mathrm{Jup}$) planets, for which the typical temperatures range from 1000 to 2000~K, and the contrasts peak in the near-IR. The downside is that these instruments are mostly insensitive to smaller and/or older planets, which are colder.

Recently, the idea of detecting Earth-like planets in the mid-IR around our closest neighbors ($\alpha$ Cen) has gained attention, and some projects aiming at doing so have been closely studied (\citeads{2017Msngr.169...16K}; \citeads{2018SPIE10702E..4AB}). In the mid-IR, the planet to star contrast ratio is maximized for planets at a few $100\,\mathrm{K}$. In principle, this wavelength range is much better suited to the detection of Earth-like planets. But the instrumental thermal background is much stronger in the mid-IR than in the near-IR, which complicates observations. Because of the longer wavelengths, the spatial resolution is also much poorer, with a loss of a factor $\approx 3-5$ in angular resolution between the mid-IR and the near-IR. But, these downsides of thermal IR observations can be mitigated by focusing on the most nearby stars whose planets would apparently be brighter and observable at greater angular separations. 

Addressing these challenges, the New Earths in the Alpha Cen Region (NEAR) experiment used the deformable secondary mirror at UT4, a new Annular Groove Phase Mask (AGPM) coronagraph optimised for the N-band and a novel internal chopper system for noise filtering \citepads{2019Msngr.178....5K} in order to improve the capabilities of VLT/VISIR in the mid-IR.

In 2019, a survey of 100 hours was carried out with VLT/VISIR-NEAR, to look for low-mass planets in the habitable zone of the $\alpha$ Cen AB binary system, a $5\,\mathrm{Gyr}$ old system located at 1.34~pc from our Sun. 
Using the best 76 hours worth of data, \citetads{2021NatCo..12..922W} reported a planet candidate C1 detected at $\mathrm{S/N} \approx 3.1 - 3.5$ in the combined data and at $\mathrm{S/N} \approx 2.5 - 2.8$  in multiple independent subsets of the data (using only the first 8 nights, the last 7 nights, or odd $/$ even nights), 
which makes it unlikely to be a random false positive although a systematic false positive cannot yet be excluded.

In this paper, we re-analyse the NEAR images with the K-Stacker algorithm (\citeads{2015tyge.conf...59L}, \citeads{2018A&A...615A.144N}, \citeads{2020A&A...639A.113L}), which looks for hidden sources following Keplerian motions in series of high-contrast images. 

Section \ref{section_search_C1_KS} presents the data set used, the K-Stacker algorithm and the main characteristics of the search carried-out. In Section \ref{section_orbital_parameters}, we report on the detection of a possible planet candidate with K-Stacker, which closely corresponds to the C1 candidate also reported in \cite{2021NatCo..12..922W}. We also discuss in this section the gain brought by K-Stacker compared to a simple co-addition of the images. 
In Section \ref{section:K-MESS} we give the completeness map obtained from the non-detection of other sources in our K-Stacker search for companions around $\alpha$ Cen A.
In Section~\ref{k_stacker_observing_strategy}, we show that K-Stacker would have been capable of recombining a series of images in which the planet would have moved significantly between observations. We show that this opens up the possibility of using a different observing strategy for such long programs, in which the observations would be split over a longer period of time, and we argue that this strategy should be followed in the future, notably to schedule the observations only under excellent weather conditions. Finally, our conclusions are given at Section \ref{conclusion}.

\begin{figure}
\centering
\includegraphics[width=\linewidth, trim=0cm 0cm 0cm 0cm, clip=True]{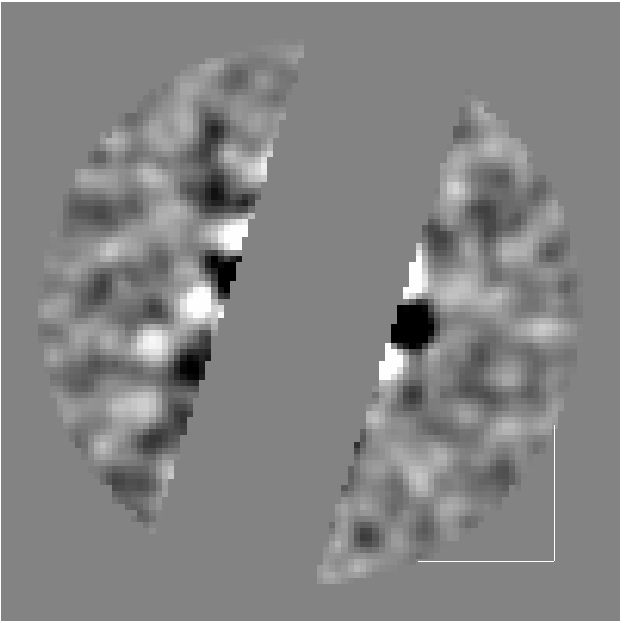}
\put(-60,13){\textbf{\textcolor{white}{1 as}}}
\put(-33,87){\textbf{\textcolor{white}{N}}}
\put(-95,23){\textbf{\textcolor{white}{E}}}
\put(-95,23){\textbf{\textcolor{white}{E}}}
\caption{First (2019-05-24) of the 16 VLT/VISIR NEAR observations used to search for a companion to $\alpha$ Cen A with K-Stacker. A diagonal strip covering approximately 36\% of the field-of-view is masked out to remove defects due to detector persistence.}
\label{fig:example_image}
\end{figure}
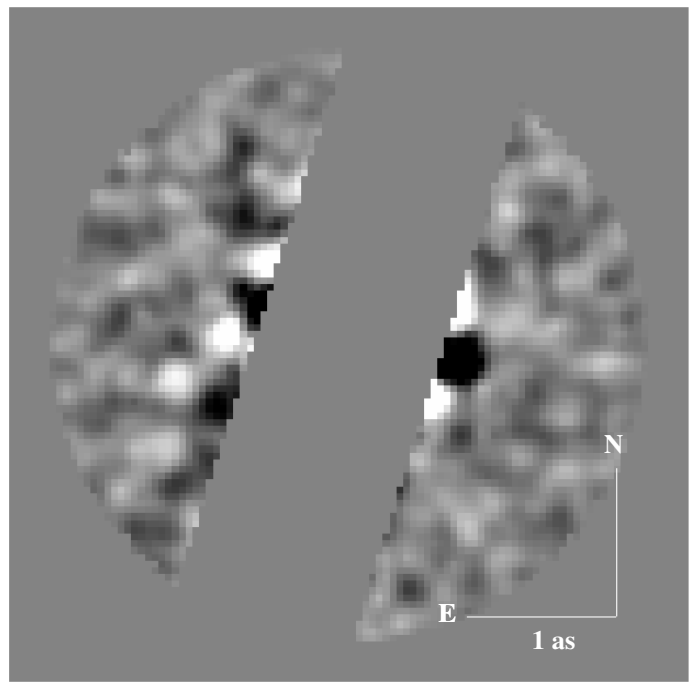

\section{Observations and K-Stacker configuration}\label{section_search_C1_KS}

\subsection{Observations}

In this work, we use observations presented in \cite{2021NatCo..12..922W} of $\alpha$ Cen, acquired using VLT/VISIR-NEAR in N-band filter (central wavelength of $11.25\, \mu$m) over a period ranging from 2019, May 23rd to 2019, June 27th. Adaptive Optics and a vortex coronagraph were used to subtract most of the starlight.
The final data products provided by the NEAR team and used by K-Stacker are a set of reduced images at 16 epochs, including bracketing pair-wise chop subtraction between $\alpha$ Cen A and B, systematic artifact removal (detector persistence and filter ghosts), a KLIP-ADI algorithm, and high-pass filtering. The final reduced images were masked along a diagonal strip to hide detector persistence residuals accumulated during the chopping sequence and not fully captured by the artifact model (see \citeads{2021NatCo..12..922W} for further details on the data processing).
Following \cite{2021NatCo..12..922W}, we rejected the observation made on May, 29th, which was of significantly lesser quality. All information about these observations and data reduction can be found in the original paper (see the main article and Supplementary material in \citeads{2021NatCo..12..922W}). An example of a fully reduced and masked image is given in Figure~\ref{fig:example_image}.


\subsection{The K-Stacker algorithm}

\noindent At its core, K-Stacker is an algorithm that looks for the maximum value of the signal to noise ratio ($\mathrm{S/N}_\mathrm{KS}$ as a function of orbital parameters). For each set of orbital parameters tested, the algorithm determines the expected positions $x_1, \dots{}, x_{16}$ of the planet at each epoch, and integrates the flux at the corresponding position in each image. This gives the signal in each image: $s_1, \dots{}, s_{16}$. In parallel, the algorithm also determines the noise levels $n_1, \dots{}, n_{16}$ at the same separations (the noise is assumed to be centro-symmetric in each image) by computing the standard deviation of a set values obtained by integrating the flux in 1000 random circles of diameter equal to one FWHM-PSF.

The $\mathrm{S/N}$ ratio is then defined as:

\begin{equation}
(\mathrm{S/N})_\mathrm{KS} = \frac{s_1 + s_2 + \dots{} + s_{16}}{\sqrt{{n_1}^2+{n_2}^2+\dots{}+{n_{16}}^2}}
\label{eq:snks}
\end{equation}

Out of all the orbits tested by the algorithm on the grid of orbital parameters, the best 100 orbits are re-optimized using Large-scale Bound-constrained Gradient (L-BFGS-B) optimization algorithm \citep{741068676f0d4748ba518263a9ca1363}. After this step, those orbits which yield a $(\mathrm{S/N})_\mathrm{KS}$ larger than a given threshold $(\mathrm{S/N})_\mathrm{thresh}$ are considered as detections of potential planets.

We can note that Eq. \ref{eq:snks} does not take into account any correction for small sample statistics, as described in \citetads{2014ApJ...792...97M}, and the quadratic sum of the individual noise terms in the denominator is a valid estimate of the total noise only in the case of uncorrelated noise between the different images of the series. Therefore, this $(S/N)_\mathrm{KS}$ quantity should be regarded only as the "gain function" optimized by our algorihtm, and any interpretation in terms of signal-to-noise should be taken with caution. In particular, we draw the attention of the reader on the fact that this $(S/N)_\mathrm{KS}$ cannot be directly compared to the $\mathrm{S/N}$ value found by \citetads{2021NatCo..12..922W}, which does include a Students' correction  \citepads{2014ApJ...792...97M}, and is based on an estimate of the noise in the centro-symmetric recombined image which takes into account possible correlations between the individual images.

\subsection{Search space}

\begin{table}[h]
\begin{center}  
\caption{Search space for K-Stacker run on the NEAR $\alpha$ Cen A images.}
  \begin{tabular}{lll}
    \hline
    \hline
    Parameter & Range & Distribution\\
    \hline
    Star mass ($M_{\text{star}}$) & $1.133$ M$_\odot$ & fixed value \\
    &&\\
    Star distance ($d_{\text{star}}$) & 1.34 pc & fixed value \\   
    &&\\
    Semi-major axis ($a$) & [0.95 au, 2. au] & uniform \\
    &&\\
    Eccentricity ($e$) & [0, 0.9]  & uniform \\
    &&\\
    Periapsis time ($t_0$) & [0 yr, 2.3 yr] & uniform \\
    &&\\
    $\Omega + \omega$\tablefootmark{$\star$}  & [$-\pi$, $+\pi$ ] & uniform \\
    &&\\
    Inclination ($i$) & [0, $+\pi$] & uniform \\
    &&\\
   $\Omega - \omega$\tablefootmark{$\star$} & [$-\pi$, $+\pi$] & uniform \\
    \hline`
  \end{tabular}
  \tablefoot{
    \tablefoottext{$\star$}{$\Omega$ is the longitude of ascending node, $\omega$ the argument of periapsis.} 
  }
  \label{tab:KS_search_parameters}
\end{center}  
\end{table}

As already noted in \cite{2021NatCo..12..922W}, at a separation of $\simeq{}1\,\mathrm{au}$, the orbital motion of the planet corresponding  approximately to the size of a resolution element cannot be neglected over the time spanned by the observations. For a circular orbit, the orbital velocity is given by:

\begin{equation}
v_\mathrm{orb} = v_\mathrm{Earth}\sqrt{M_\mathrm{star}/a}
\end{equation}
\noindent{}where $M_\mathrm{star}$ is the mass of the star expressed in Solar masses, $a$ the semi major-axis expressed in au, and $v_\mathrm{Earth}=29.8\,\mathrm{km/s}$ is the Earth's orbital velocity.

For a face-on circular $a=1\,\mathrm{au}$ orbit around $\alpha$ Cen A (a $1.1\,M_\odot$ star at 1.34 pc) this corresponds to a projected angular velocity of 12.8 mas/day, and a total orbital motion of more than 350 mas between the first and last epochs, comparable to the Airy pattern's full width at half maximum. Such displacement can be taken into account by a K-Stacker algorithm \citepads{2020A&A...639A.113L} in order to find the best possible way to combine all epochs accounting for the orbital motion of a putative planet.

We configured K-Stacker to search for planets over the entire field-of-view of the reduced NEAR images (1.5 $''$ in radius), with the persistent diagonal strip masked (see Fig. \ref{fig:example_image}). This corresponds to projected separations ranging from 0.95~au to 2~au. We did not constrain the eccentricity, and explored the range 0.0-0.9. All other parameters are explored over their entire range of possible values, with the notable exception of the stellar parameters (mass and distance), which are taken as fixed values. The number of steps in the grid of parameters explored by the brute-force stage of K-Stacker was determined using an empirical method described in \cite{2018A&A...615A.144N}. A total of $3.5\times{}10^5$ orbits with companions positions outside the masks (at the epochs of observation), covering the range of parameters given in Table~\ref{tab:KS_search_parameters} were set.

\section{Recovery of the C1 planet candidate}
\label{section_orbital_parameters}

\begin{figure}[h]
\begin{center}
   \includegraphics[width=1. \linewidth, clip=True]{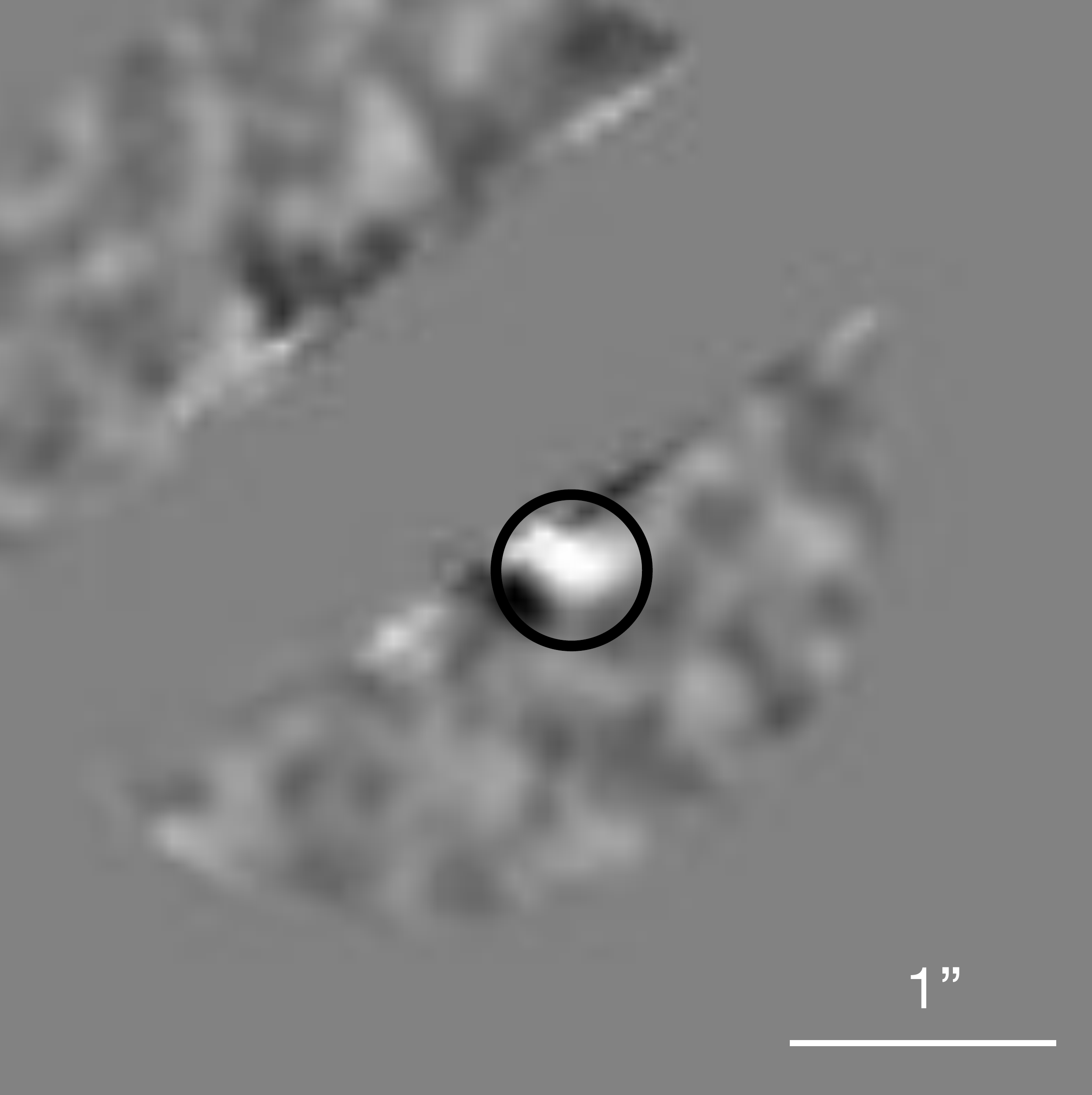}
   \caption{Alpha Cen A NEAR images recombined by using the best orbital parameters recovered by K-Stacker ($a=0.96\,\mathrm{au}$, $e=0.78$, $t0=-0.11\, \mathrm{yr}$, $\Omega=5.34\, \mathrm{rad}$, $i=1.052\, \mathrm{rad}$, $\theta_0=0.54\, \mathrm{rad}$). At each epoch, the images are rotated and shifted to put C1 on its periastron position found by K-Stacker, and the frames are co-added. The candidate planet is detected at a $\snks$ level of 5.5.}
     \label{fig:ImageKS_C1}
     \end{center}   
\end{figure}

\begin{figure*}[!h]
\includegraphics[width=18.3cm]{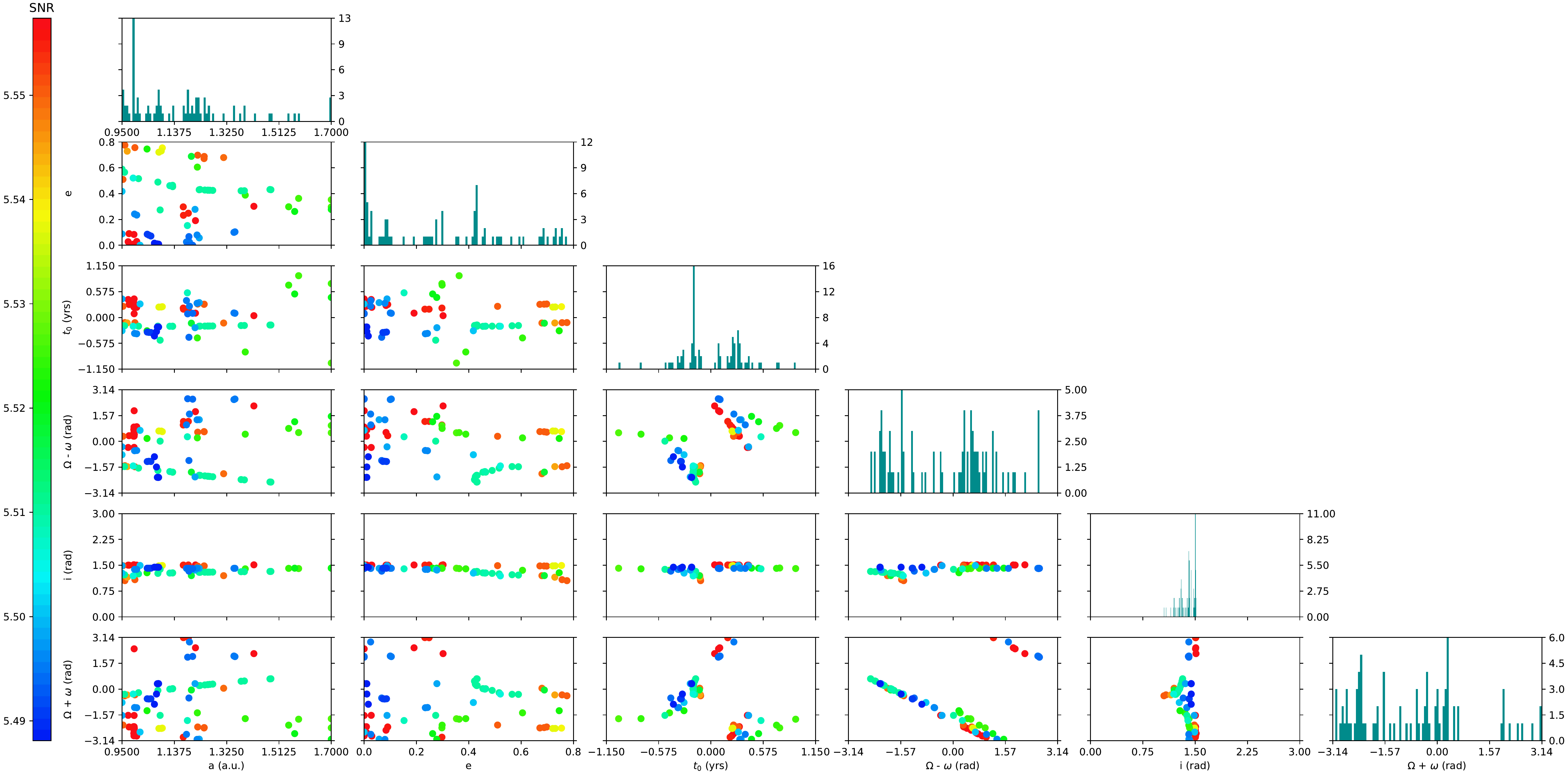}
\caption{Corner plots of the orbital parameters found by K-Stacker for $\alpha$ Cen A C1. At left and bottom: scale of the orbital parameters. At right: scale of the histograms. The 84 points showed in each 2D diagram corresponds to the 84 orbits with the highest $\snks$ found by K-Stacker. The color of each point gives the K-Stacker signal to noise indicated at left. The origin of the t$_0$ K-Stacker date is  $2019-05-24$.} 
\label{fig:coner_plot}
\end{figure*}

\begin{figure*}[h!]
\begin{center}
   \centering
\includegraphics[width=1. \linewidth, clip=True]{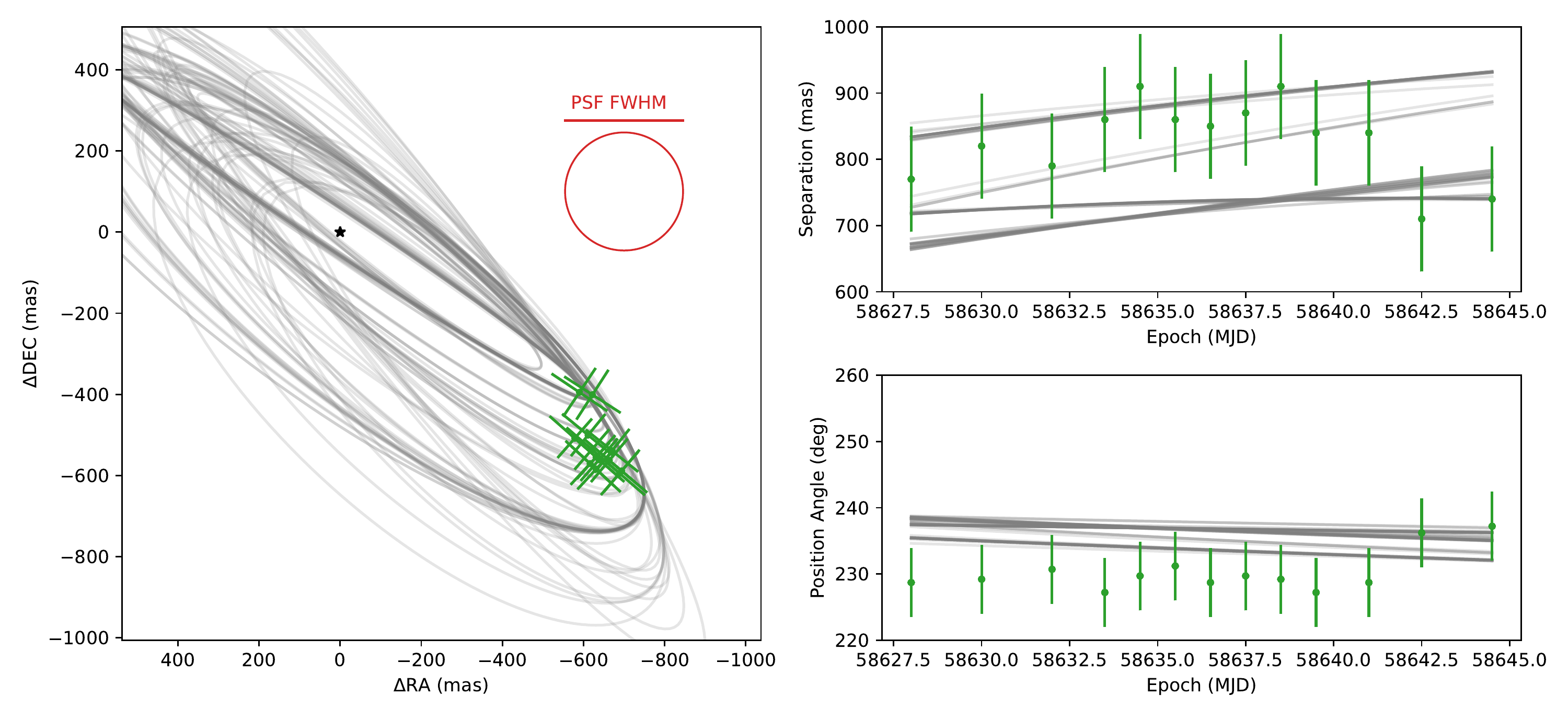}
   \caption{Left: plot at two dimensions of the orbits projected on the sky.
   The 84 best orbits found by K-Stacker are shown in grey. 
   The green crosses show the astrometric positions with their error bar given by \citetads{2021NatCo..12..922W}. The red circle shows the size of the FWHM-PSF used to integrate the flux in K-Stacker. Right: zoom at the positions found by K-Stacker and \citetads{2021NatCo..12..922W} in one dimension plots in separation and position angles.}
     \label{fig:ks_orbits}
     \end{center}   
\end{figure*}

Over the $3.5\times{}10^5$  orbits tested, K-Stacker found $100$ orbits that all yield $\snks > 5.4$  in the recombined image. 
Following the same method presented in Figure 2 of \citetads{2020A&A...639A.113L}, we keep the 84 best orbits corresponding to a relatively constant level of $\snks \approx 5.5$, before a first strong decrease. The corresponding combined image is given in Figure~\ref{fig:ImageKS_C1}.

Although, they have different orbital parameters, typically ranging from $a=0.95\,\mathrm{au}$ to $a=1.32\,\mathrm{au}$ and $e$ spread over the full range of 0 to 0.9 (see corner plots in Figure~\ref{fig:coner_plot}), all these orbits pass by similar positions at the epochs of observations (see grey curves in Figure~\ref{fig:ks_orbits}). This is evidence of the repeated presence of a significant feature in the data set, whose motion is compatible with Kepler's laws. Interestingly, the positions at all epochs along these best 84 orbits found by K-Stacker are not only similar between different orbits, but also closely match the astrometric positions derived for the C1 candidate by \citetads{2021NatCo..12..922W}. The K-Stacker orbits are slightly to the right of the  astrometric points \cite{2021NatCo..12..922W} at less than 1/4th of the instrument PSF of $\approx 300\, \mathrm{mas}$ (see grey curves and green crosses in the zoom of Fig. \ref{fig:ks_orbits}). This small systematic discrepancy may be due to the very faint signal  ($\mathrm{S/N} \approx 1-2$) in the \cite{2021NatCo..12..922W} images to extract the astrometry of C1. Their measurements may have been biased toward the stripes artifact parallel to the chopping direction while K-Stacker is more sensitive to the part of the signal that moves (i.e. less affected by this systematic). Moreover, \citet{2021NatCo..12..922W} have extracted the astrometric positions of C1 by injection of a negative PSF in groups of two nights (i.e. if the astrometry is shifted at one epoch, it will be also shifted at the next epoch). Finally, most of the K-Stacker orbits have their separations that increase with the epochs but Fig. \ref{fig:ks_orbits} shows that a few possible K-Stacker orbits have separations that increase and then decrease, in accordance to the astrometric solutions given by \citeads{2021NatCo..12..922W} (See Supplementary Fig. 5) where the separation decreases in the last two epochs.

\begin{figure}[!h]
\centering
\includegraphics[width=1. \linewidth, clip=True, trim=0.8cm 0cm 1.5cm 1cm]{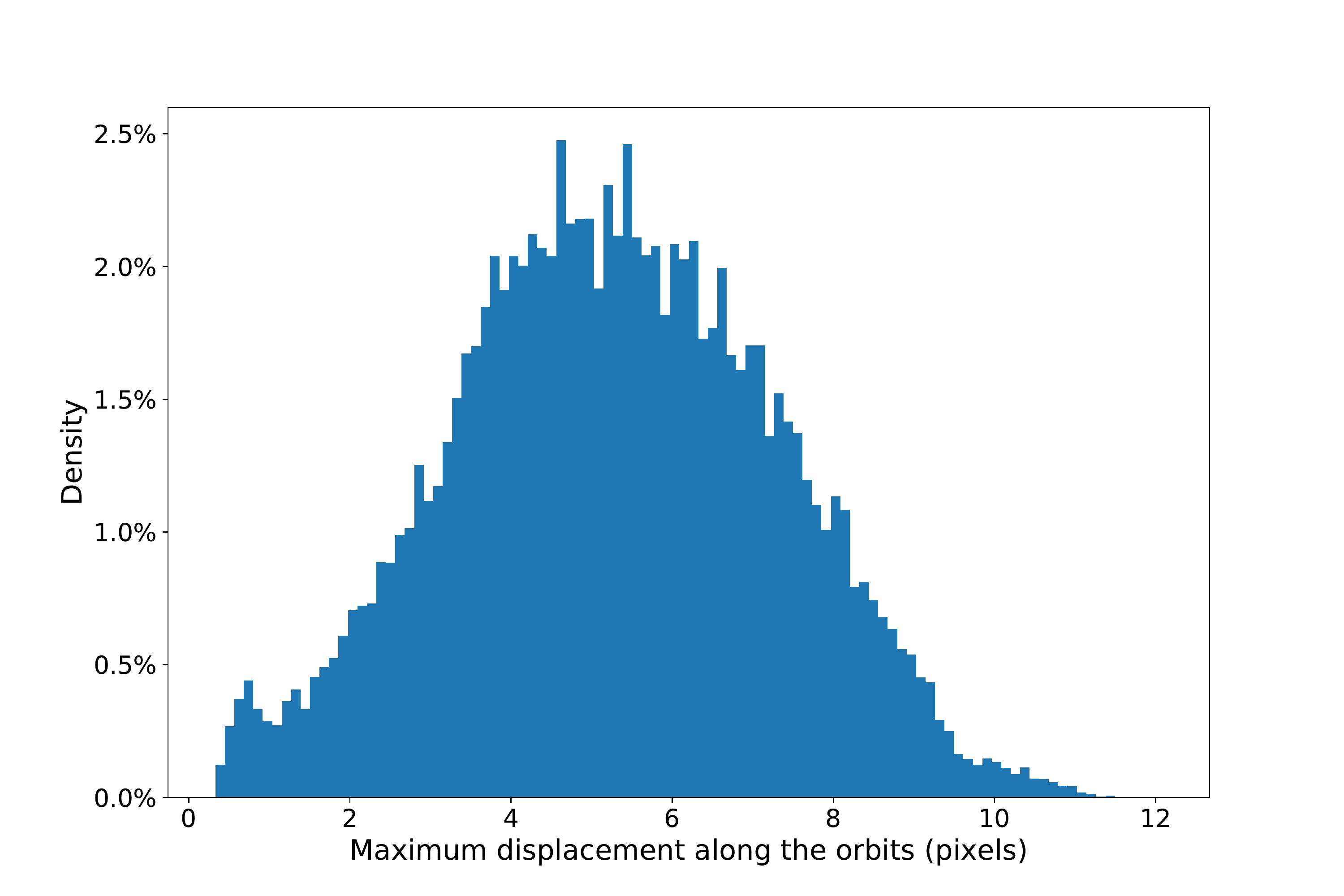}
\caption{Distribution of the maximum displacement of searched planets along the orbits tested by K-Stacker for the orbital parameters ranges shown at Table \ref{tab:KS_search_parameters}.} 
\label{fig:histo_max_dist_orbits}
\end{figure} 

\begin{figure}[!h]
 \begin{center}
\includegraphics[width=1. \linewidth, clip=True]{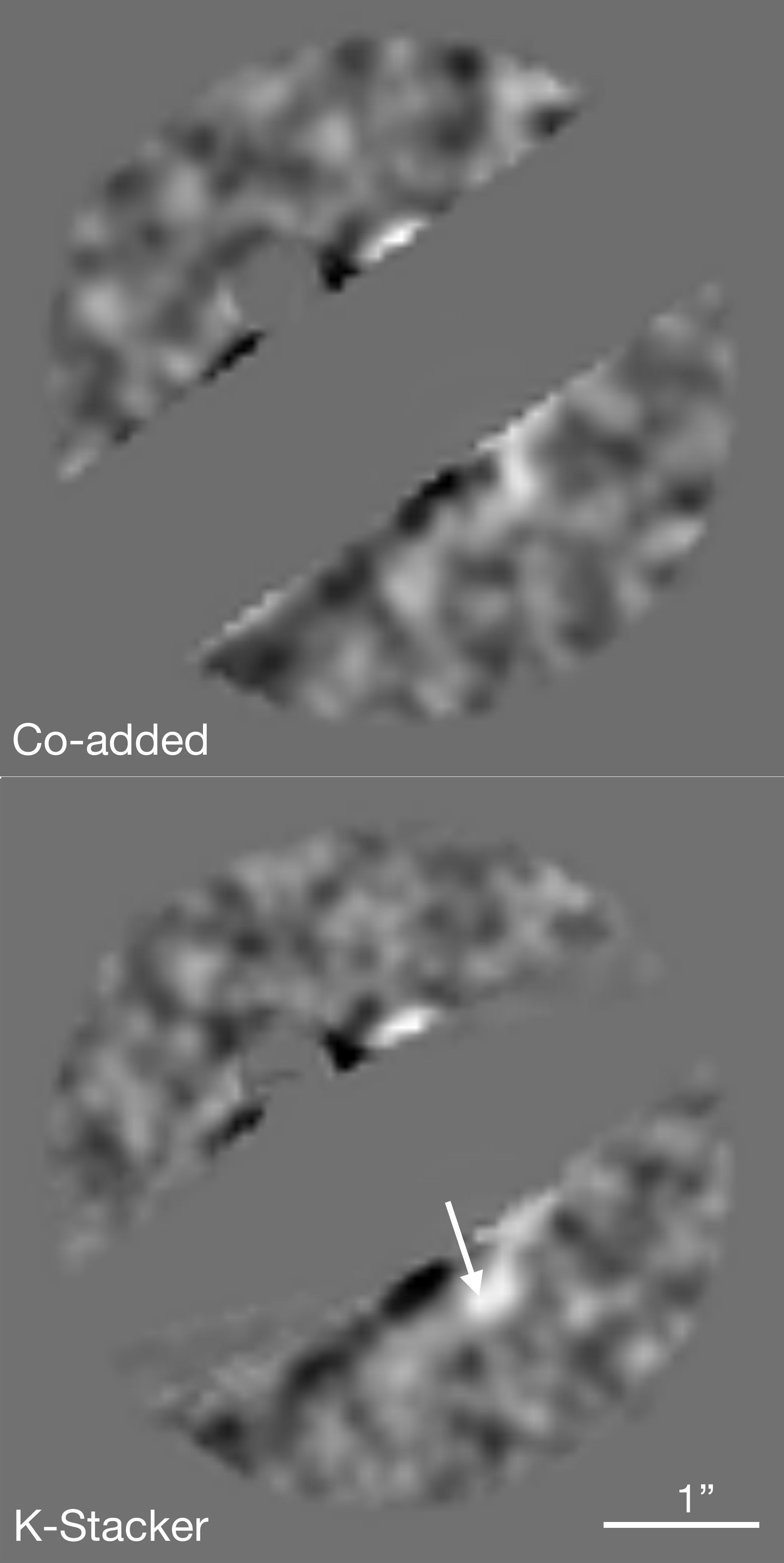}
\caption{Top: Sum of the 16 reduced images of \citeads{2021NatCo..12..922W} where a fake planet moving of $\approx 9$ pixels along one of the orbit ($a=1.58\,\mathrm{au}$, $e=0.13$, $t0=-1.78\, \mathrm{yr}$, $\Omega=-3.14\, \mathrm{rad}$, $i=0.38\, \mathrm{rad}$, $\theta_0=2.2\, \mathrm{rad}$) tested by K-Stacker (see Fig. \ref{fig:histo_max_dist_orbits}) has been injected at $\mathrm{S/N} \approx 1$. C1 has been masked by a small circle visible at top-left. The planet is not detectable with a simple co-addition of the images. Bottom: The same images have been recombined by K-Stacker. 
We detect the planet at the position shown by the arrow at an $\snks \approx 5$. } 
\label{fig:recombine_orbit_9_pixel}
 \end{center} 
\end{figure} 

The K-Stacker orbits (Fig. \ref{fig:coner_plot}) are very similar to the orbital solution derived by \citetads{2021NatCo..12..922W} from their astrometric solutions using a Keplerian model fit \citepads{2020AJ....159...89B}. Both methods find a semi-major axis $a \approx 1$ au with an inclination of $i \approx 60-80$ deg, similar to the inclination of the binary's orbit which is 79$^\circ$ \citep{2016A&A...594A.107K}. The other orbital parameters are not well constrained, due to the limited motion ($\approx$ 1 PSF) over the range of epochs spanned by the observations (see also discussion in Sect. \ref{k_stacker_observing_strategy}). 
Overall, the fact that K-Stacker reaches a $\mathrm{S/N} > 5.4$ over 84 different orbits which all corresponds to positions similar to the one derived by \cite{2021NatCo..12..922W} for their C1 candidate lead us to conclude that K-Stacker has detected the same candidate. In accordance to \citetads{2021NatCo..12..922W} study, the 84 best orbits found by K-Stacker give all a maximum displacement smaller than $6$ pixels during the period of observation. Within the error bars, there is no significant gain in terms of $\mathrm{S/N}$ obtained using K-Stacker over a simple co-addition of the images, as used in \citeads{2021NatCo..12..922W}. Given the limited motion of the candidate over the range of epochs, K-Stacker is also unable to unambiguously distinguish a true planet candidate from, for example, a disk. When masking the candidate in all images, and injecting an inclined disk model of similar brightness to C1 (see Supplementary material in \citeads{2021NatCo..12..922W}) at a different position angle in the images, K-Stacker also tends to converge towards a solution with $\mathrm{S/N}>5.0$, confusing the edge of a disk with a true planet candidate. 

Fig. \ref{fig:histo_max_dist_orbits} shows the distribution of the maximum displacement along the orbits tested by K-Stacker that peaks at $\approx 5$ pixels ($\approx$ half a PSF) but the tail of the distribution reaches a displacement of 11 pixels. Assuming a linear motion of the planet with time, adopting a series of 16 epochs spaced in a similar way as the observations taken by \citetads{2021NatCo..12..922W}, and using the post-ADI PSF structure of the VLT/VISIR observations, we find that the loss in $\snks$ (without accounting for this motion by computing the noise and signal always at the middle epoch) reaches $\simeq{}15\%$ for a total displacement of 7 pixels, and 25\%  that can not be ignored for a displacement of 11 pixels.
 
This has two consequences: (1) compared to the co-addition performed by \cite{2021NatCo..12..922W}, our K-Stacker based approach provides a $\snks$ boost of $15\%$, for a large number of possible orbits ($\simeq{}30\%$), and up to $25\%$ in some cases, which could potentially have led to new detection in the field of view of the NEAR-VISIR reduced images (see Fig~\ref{fig:recombine_orbit_9_pixel}, in which we provide an example of the impact of K-Stacker in such a case); (2) the loss in S/N induced by the orbital motion should be accounted for in the calculation of the detection limits.
 
 \section{Search completeness analysis} \label{section:K-MESS}

Each term $s_k$ in Equation~\ref{eq:snks} is the sum of two components: the signal arising from speckles and other sources of noise $q$, and a potential planet contribution $p$. Considering an orbit tested by K-Stacker which results in a given value of $(\mathrm{S/N})_\mathrm{KS}$ and assuming that no planet is present in the images on this orbit, the measured $(\mathrm{S/N})_\mathrm{KS}$ corresponds only to the speckle and noise contributions. In that case, an additional planet would be detected  on this orbit if its integrated fluxes in all images $f_1, \dots{}, f_{16}$ was such that:

\begin{equation}
f_1+\dots{}+f_{16} = \left[(\mathrm{S/N})_\mathrm{thresh} - (\mathrm{S/N})_\mathrm{KS}\right]\times{}\sqrt{{n_1}^2+{n_2}^2+\dots{}+{n_{16}}^2}
\end{equation}

To relate the $f_1, \dots{}, f_{16}$ terms to the true planet to star contrast, the images need to be flux calibrated. In order to do so, we injected a planet at a contrast ratio of $10^{-4}$ in all raw images, and measured the reference planet signal $f_{\mathrm{ref}1}, \dots{}, f_{\mathrm{ref}{16}}$ after the reduction as calculated by K-Stacker at the position where the planet was injected. We performed this injection at several values of separations (ranging from 0.7 to 1.5 au) to take into account the impact of the ADI reduction on the planet flux. The minimum contrast required for the planet to be detected can then be written:

\begin{equation}
C_\mathrm{min} = \frac{\left[(\mathrm{S/N})_\mathrm{thresh} - (\mathrm{S/N})_\mathrm{KS}\right]\times{}\sqrt{{n_1}^2+\dots{}+{n_{16}}^2} }{f_{\mathrm{ref}1}+\dots{}+f_{\mathrm{ref}16}}\times{}10^{-4}
\label{eq:cmin}
\end{equation}

This equation assumes that all the flux of the planet PSF is properly integrated by K-Stacker. Since we are working with heavily masked images (see Figure~\ref{fig:example_image}), we also need to consider the case when the planet falls behind the masked area. For this, we increase the required minimum contrast $C_\mathrm{min}$ by a factor $1/f_\mathrm{out}$ where $f_\mathrm{out}$ is the fraction of the flux of the PSF actually detectable by K-Stacker (i.e. outside of all masked areas). 

We calculated the value of this minimum contrast for each of the $3.5\times{}10^5$ orbits tested by K-Stacker, and grouped the results by value of semi-major axis. This gives a set of about $6 \times{}10^4$ orbits for each of the 6 values of semi major-axis, covering a range of possible orientation in the images. For each set, we can then proceed to extract the value of the different percentiles of the distribution of $C_\mathrm{min}$, which give estimates of the probability of detecting a planet at this contrast level with K-Stacker. This method of calculating the detection limits overlooks several different problems, such as the fact that the search on the grid can underestimate the flux of the planet, or the fact that low-flux undetected planets in the images can already contribute to the calculation of the $(\mathrm{S/N})_\mathrm{KS}$ term of Equation~\ref{eq:cmin}.

To convert the contrast ratios to planetary radii, we used the same method as \cite{2021NatCo..12..922W}. For a given semi major-axis $a$, the equilibrium temperature of the planet is calculated using:

\begin{equation}
T_\mathrm{eq} = T_\mathrm{star}\sqrt{\frac{R_\mathrm{star}}{2a}} (1-A_\mathrm{B})^{1/4}
\end{equation}
where $T_\mathrm{star} = 5500\,K$, $R_\mathrm{star} = 1.2\,R_\odot$, and $A_\mathrm{B}$, the Bond albedo, is 0.3.

The temperature of the planet is set by radiation from the star, plus an extra contribution from internal heating:

\begin{equation}
T_\mathrm{planet} = T_\mathrm{eq}\times{}(1+f_\mathrm{extra})
\label{eq:Tplanet}
\end{equation}

\begin{figure}[!h]
\centering
\includegraphics[width=1. \linewidth, clip=True, trim=0.8cm 0cm 1.5cm 1cm]{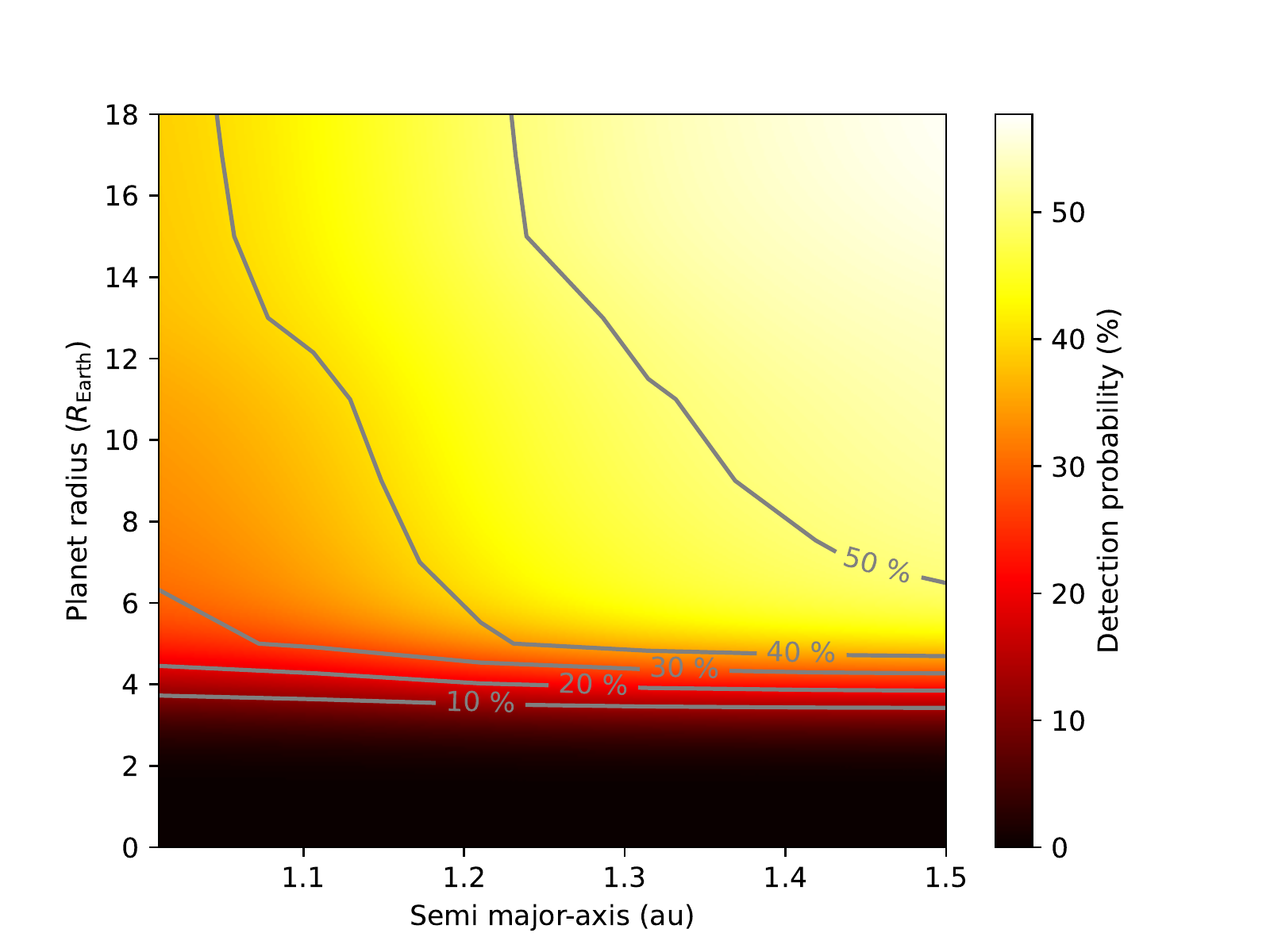}\vfill
\includegraphics[width=1. \linewidth, clip=True, trim=0.8cm 0cm 1.5cm 1cm]{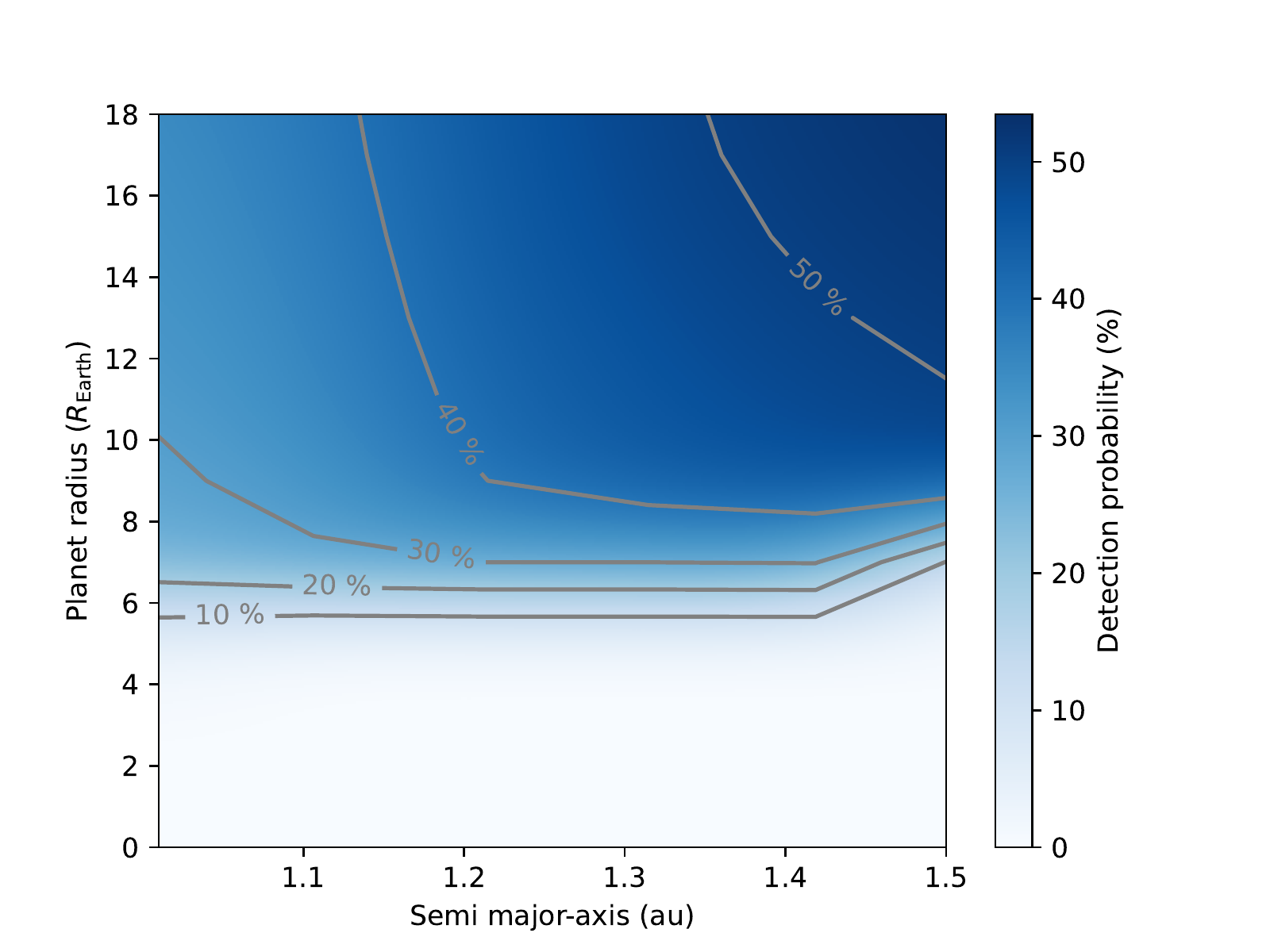}
\caption{Results of our completeness analysis showing the probability of detecting a planet of a given radius at a given semi major-axis, in the case of a hot planet (upper panel, $f_\mathrm{extra} = 0.5$ in Equation~\ref{eq:Tplanet}), and a cold planet (bottom panel, $f_\mathrm{extra} = 0.1$).}
\label{fig:messKS}
\end{figure}

To allow for a direct comparison with \cite{2021NatCo..12..922W}, we consider two cases: a hot case with $f_\mathrm{extra} = 0.5$, and a cold case with $f_\mathrm{extra} = 0.1$. The results obtained for both the cold and hot cases are presented in Figure \ref{fig:messKS}.

The completeness of our search is typically about $30 \%$ at a radius of $\simeq 8\, R_\mathrm{Earth}$, and limited to about $60 \%$ for all radii, due to the significant area impacted by defects and excluded of the search area (see Fig~\ref{fig:example_image}).

 \section{The case for a more efficient K-Stacker based observing strategy} \label{k_stacker_observing_strategy}
 
\citetads{2021NatCo..12..922W} have observed on a short period of one month in order to be able to sum the reduced images at all the epochs without needing to take into account the Keplerian motion to avoid large losses in S/N. This strategy, while it does alleviate the need for an algorithm capable of dealing with Keplerian motion in the recombination (at least in most cases, see Section~\ref{section_orbital_parameters} for a discussion on the impact of the orbital motion on the S/N), also significantly hampers our ability to determine the orbital motion of the planet, and thus its true separation from the star.

\begin{figure}
\centering
\includegraphics[width=1. \linewidth, clip=True]{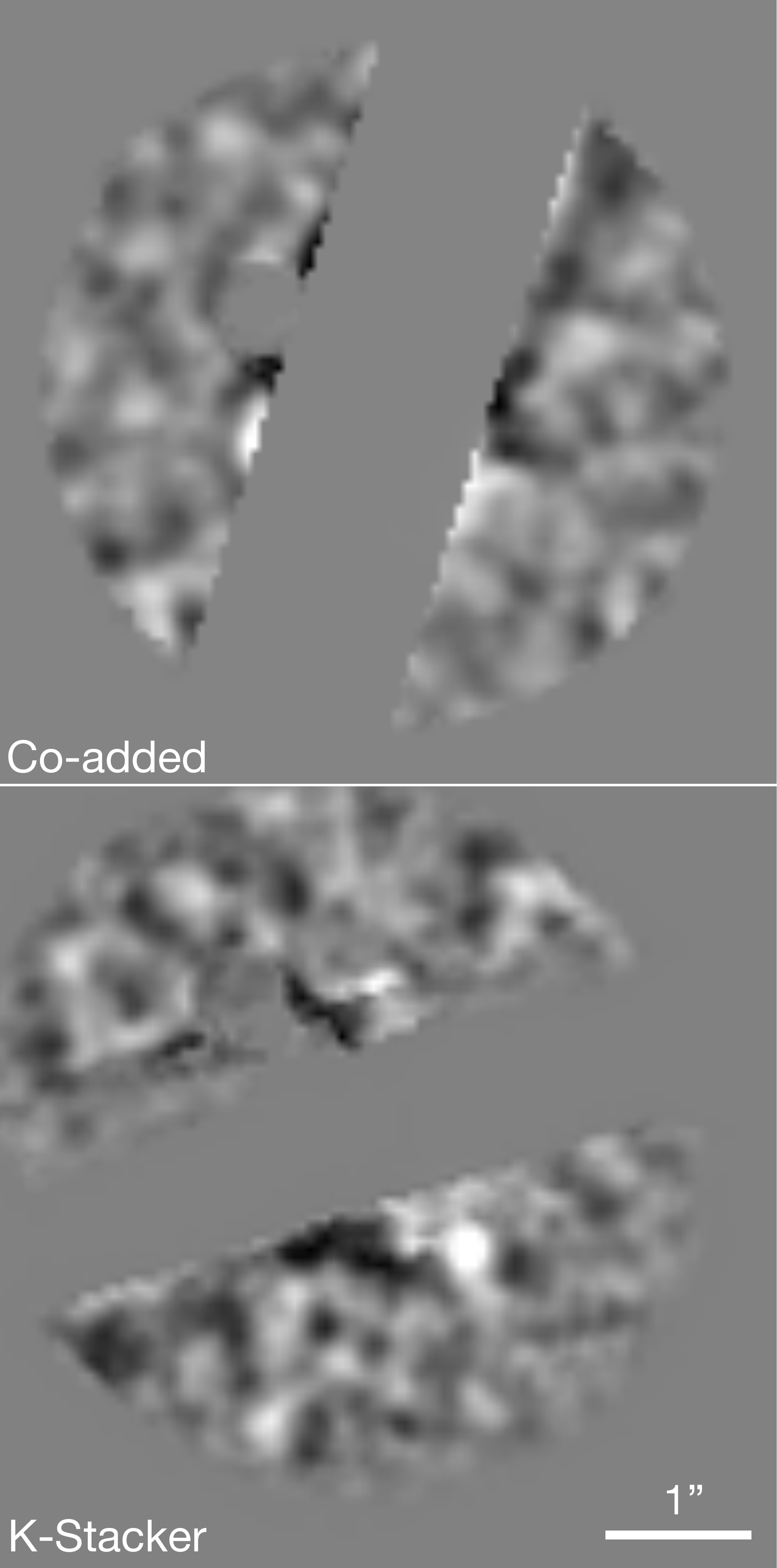}
\caption{A fake planet has been injected on the same orbit than C1, but with 0.03 year between each observation (i.e. $\approx 0.5 \mathrm{yr}$ during the 16 epochs). Top: The planet is undetectable when the images are co-added without taking into account the Keplerian motion.
Bottom: Best recombined image resulting from the K-Stacker run.  The planet is easily detected at a $\snks$ level of $\approx 5.5$.}
\label{fig:co-added_images_fake_planet_KS_orbit_05Yr}
\end{figure}

\begin{figure*}
\includegraphics[width=17cm]{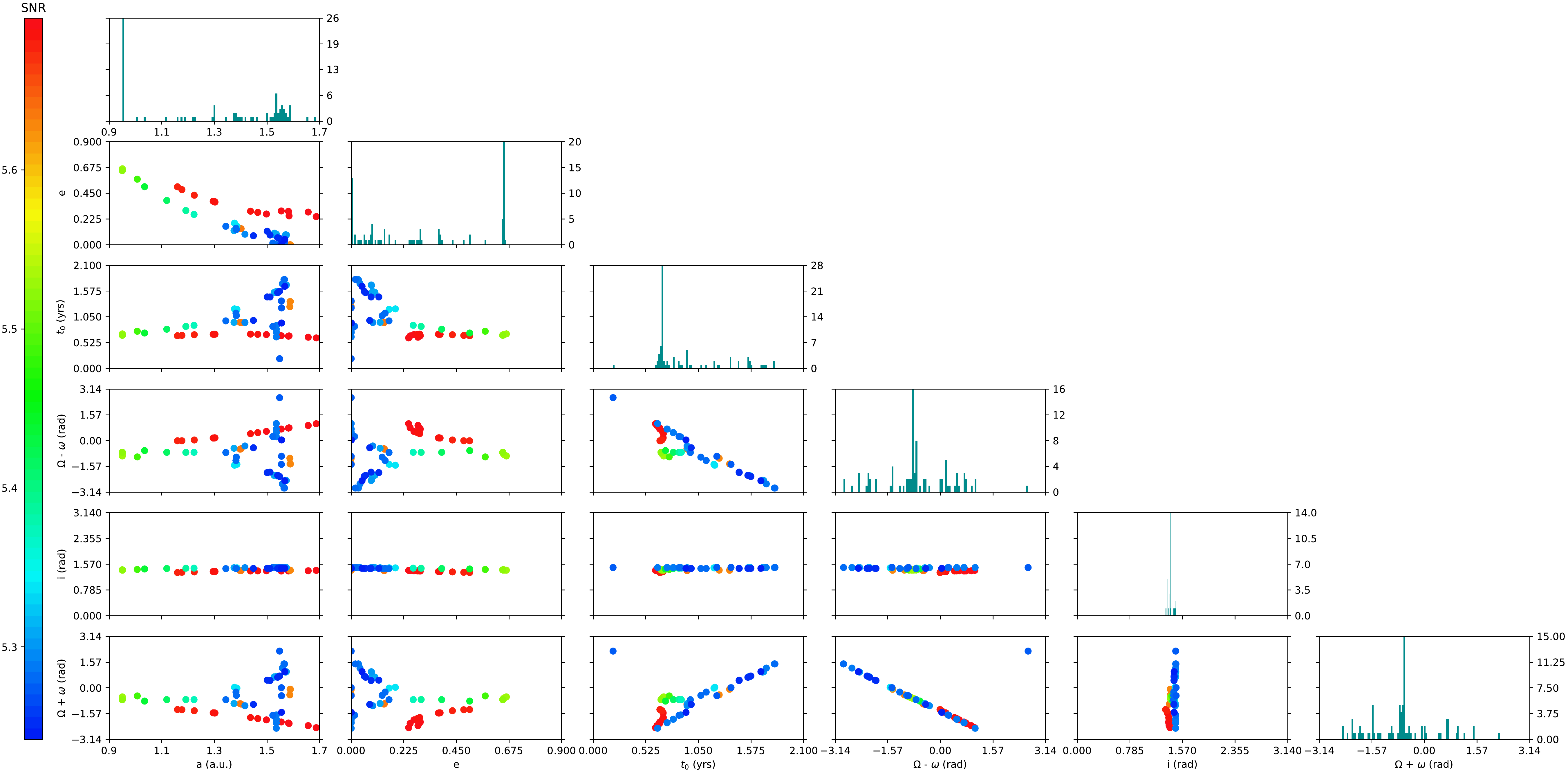}
\caption{Corner plots of the orbital parameters found on a planet injected on the best C1 K-Stacker orbit but observed during $0.5\,\mathrm{yr}$. At left, and bottom: scale of the orbital parameters. At right: scale of the histograms. The 84 points showed in each 2D diagram corresponds to the 84 orbits with the highest $\snks$ found by K-Stacker. The color of each point gives the K-Stacker $\snks$ indicated at left.} 
\label{fig:coner_plot_0.5Yr_mean_KS}
\end{figure*}

Given the ability of K-Stacker to recombine images while properly taking into account the orbital motion, we decided to simulate an hypothetical scenario, in which the NEAR observations would have been acquired over a few months, covering a significant part of the expected orbital period (i.e. $\sqrt{a^3 / m_{star}}\, \approx\, 1 \, \mathrm{yr}$ a planet at about one $\mathrm{au}$ from $\alpha$ Cen A). The NEAR data set was recorded over three weeks in quite variable conditions, so the data should be representative also for a campaign of several months. The goal was to determine whether or not K-Stacker would have been capable of properly recovering the orbital parameters in such a scenario. To test this scenario, we rotate the NEAR / VISIR reduced images by 180 degrees, and inject a fake planet in the reduced images at the same S/N $\approx 1$ in each image (i.e. equivalent to injecting a companion at a location of $180 \deg$ from the C1 position), using the best orbital parameters recovered by K-Stacker for the C1 candidate ($a=0.96\,\mathrm{au}$, $e=0.78$, $t0=-0.11\, \mathrm{yr}$, $\Omega=5.34\, \mathrm{rad}$, $i=1.052\, \mathrm{rad}$, $\theta_0=0.54\, \mathrm{rad}$). Assuming that the observations were obtained regularly (i.e. one image every 11 days) over a period of 6 months (about half of the orbital period), the planet drifts significantly in the images, and a naive co-addition results in a non-detection (see Fig.~\ref{fig:co-added_images_fake_planet_KS_orbit_05Yr}). However, when the images are combined with K-Stacker, the planet is detected at a S/N level of 5.5, similar to the S/N obtained on the true C1 candidate (see Fig.~\ref{fig:co-added_images_fake_planet_KS_orbit_05Yr}). The distribution of the most probable orbital parameters resulting from this recombination are given in Fig.~\ref{fig:coner_plot_0.5Yr_mean_KS}, and shows that the orbital parameters are overall well constrained, with a strong peak in the distribution around $a=0.95 \,\mathrm{au}$, $e=0.7$, $t0=-0.18 \, \mathrm{yr}$, $i=1.4 \, \mathrm{rad}$, $\Omega = 5.5 \, \mathrm{rad}$, $\theta_0 = 0.1 \, \mathrm{rad}$. These results should be compared with Fig. \ref{fig:coner_plot}, which gives the contraints derived from the observations actually obtained. Allthough the total observing time is the same in both cases, the orbital parameter estimates are much better when splitting the observations and relying on K-Stacker to perform the recombination properly. In appendix \ref{appendix}, we also provide a second example of the same scenario, using a different set of orbital parameters. The conclusions are unchanged.

\section{Discussion and conclusion} 
\label{conclusion}

 K-Stacker has detected a point source at the same position as the C1 candidate found by \citetads{2021NatCo..12..922W}. The $\snks$ is equal to $5.5$ in the recombined image, corresponding to a planet of $\approx 4$ and $\approx 7$ Earth radius respectively for the hot and cold models ($f_\mathrm{extra} = 0.5$ or 0.1 in Equation~\ref{eq:Tplanet}). This is in very good agreement with the values reported by \citetads{2021NatCo..12..922W}. \\
We showed that for $\approx 15 \%$ of the orbits tested by K-Stacker, the inclusion of the orbital motion in the recombination of the images provided a boost of $\simeq{}20-25\%$ in terms of S/N. But despite this boost, we report no new detection in this dataset. This study demonstrates for the first time in mid-infrared high contrast imaging that K-Stacker is able to detect hidden planets (here at an $\mathrm{S/N} \approx 1$ in each image) in a series of images by increasing significantly the signal to noise ratio. This new K-Stacker validation is fundamental for future observations. Indeed, \citetads{2021NatCo..12..922W} have been able to detect C1 by summing several observations without taking into account the Keplerian motion smaller than $1$ PSF in the mid-infrared during their period of $\approx 3$ weeks of observations. However, this will not be always possible. For example, the detection of C1 in the visible with an instrument like SPHERE-Zimpol would take about $70$ hours. At visible wavelengths, the orbital motion will be much larger than one PSF over a minimum of $\approx 30-50$ nights required to reach this amount of observing time. So, an algorithm such as K-Stacker becomes a key component for such an observation. K-Stacker could also help to search for Jupiter at 3-10 au around the nearest young stars, with first generation E-ELT instruments like MICADO and HARMONI. Theoretically, an earth like planet will be detectable without K-Stacker, in $5-10$ hours with the future mid-IR instrument METIS at the E-ELT focus. But, if nothing is detected, K-Stacker will become absolutely necessary to reach the higher contrast with long exposure time under the best atmospheric conditions (i.e. for observations spread over more than 10 days with METIS).

This could be an opportunity to revise our direct-imaging observing strategy: instead of concentrating all the data around a single epoch to allow for a simple stacking of the images, we recommend that the observations be split around several epochs, covering as much of the orbital period as possible, and then recombined using K-Stacker to better constrain the orbital parameters. This K-Stacker strategy will also allow scheduling observations only under excellent weather conditions, in order to reach the ultimate contrast that the instruments can provide. The K-Stacker code is available on gitlab.lam (https://gitlab.lam.fr/RHCI/kstacker) and github (https://github.com/kstacker). It can be used for both preparation and exploitation of direct imaging observations with current and future planet imagers. 

\begin{acknowledgements}

NEAR was made possible by contributions from the Breakthrough Watch program, as well as contributions from the European Southern Observatory, including director’s discretionary time. Breakthrough Watch is managed by the Breakthrough Initiatives, sponsored by the Breakthrough Prize Foundation. K.W. acknowledges support from NASA through the NASA Hubble Fellowship grant HST-HF2-51472.001-A awarded by the Space Telescope Science Institute, which is operated by the Association of Universities for Research in Astronomy, Incorporated, under NASA contract NAS5-26555. This work was supported by the Programme National de Planetologie (PNP) of CNRS-INSU co-funded by CNES. This research has made use of computing facilities operated by CeSAM data center at LAM, Marseille, France.

\end{acknowledgements}

%
%
\bibliographystyle{aa} 
\bibliography{lecoroller_AlphacenA_biblio}   

\begin{appendix} 

\onecolumn

\section{K-Stacker observing strategy gain} \label{appendix}

In this second example of a hypothetical scenario in which the NEAR images would have been acquired with a different strategy (see Section~\ref{k_stacker_observing_strategy}), we have inject a fake planet in the $\alpha$ CenA images (rotated by $180$ deg and with C1 masked) along another possible orbit found by K-Stacker (see Fig. \ref{fig:coner_plot} \& \ref{fig:ks_orbits}). This orbit corresponds to a larger semi-major axis and a moderate eccentricity (i.e. $a=1.55\,\mathrm{au}$, $e=0.3$, $t_0=-1.09\, \mathrm{yr}$, $\Omega=2.44\, \mathrm{rad}$, $i=1.41\, \mathrm{rad}$, $\theta_0=1.65\, \mathrm{rad}$).

\begin{figure*}[!h]
\includegraphics[width=16 cm]{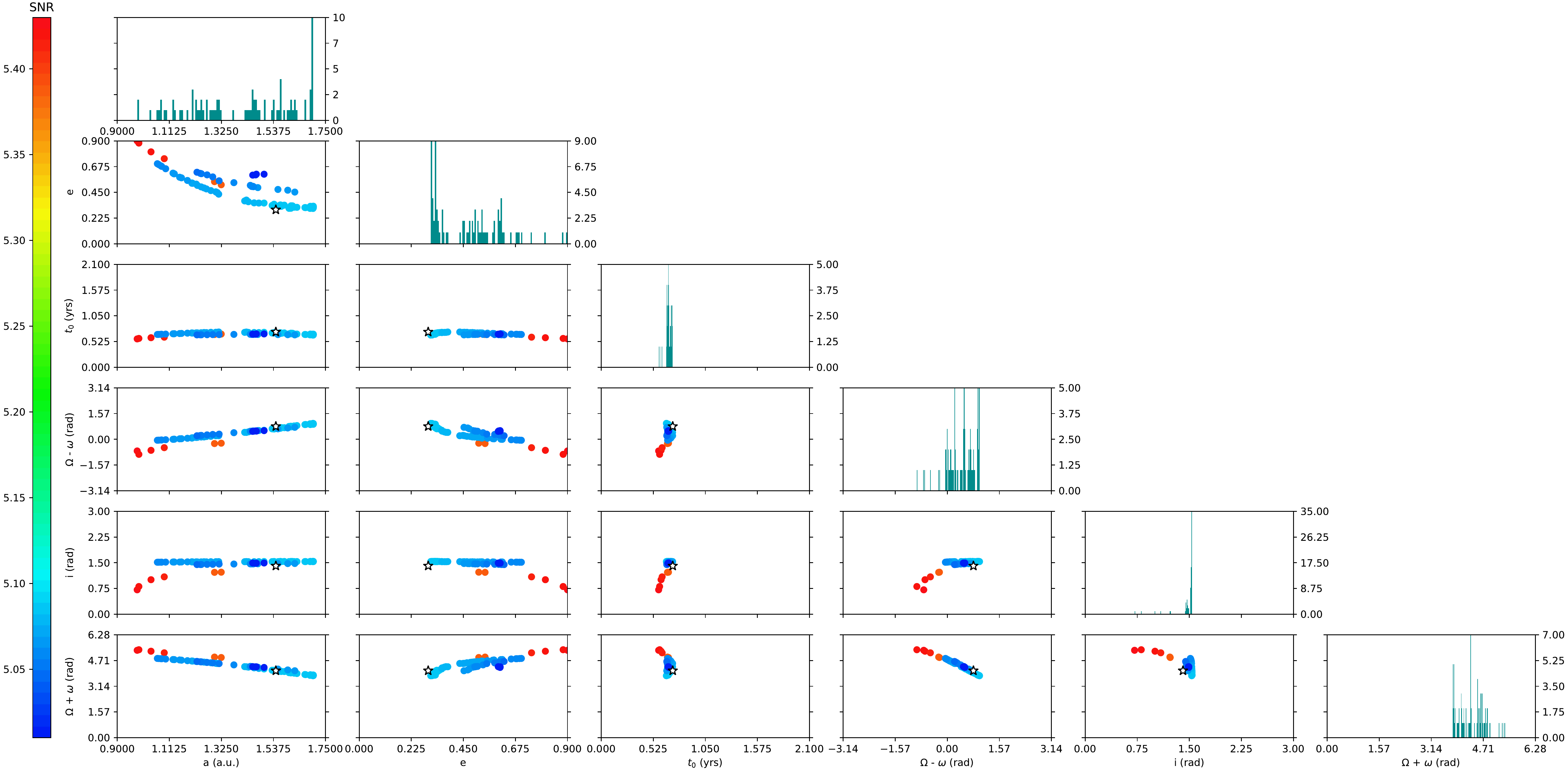}
\caption{Corner plots of the orbital parameters found by K-Stacker in a hypothetical scenario in which the NEAR images span a total of 6 months (one observation every 11 days), and assuming that the planet is following an orbit given by $a=1.55\,\mathrm{au}$, $e=0.3$, $t0=0.72\, \mathrm{yr}$, $\Omega=2.44\, \mathrm{rad}$, $i=1.41\, \mathrm{rad}$, $\theta_0=1.65\, \mathrm{rad}$.
The 84 points showed in each 2D diagram corresponds to the 84 orbits with the highest $\snks$ found by K-Stacker. The color of each point gives the K-Stacker signal to noise indicated at left. The star symbol shows the position of the injected planet in the corner plots.} 
\label{fig:coner_plot_0.5Yr_orbit15}
\end{figure*}

\begin{figure*}[!h]
\includegraphics[width=16 cm]{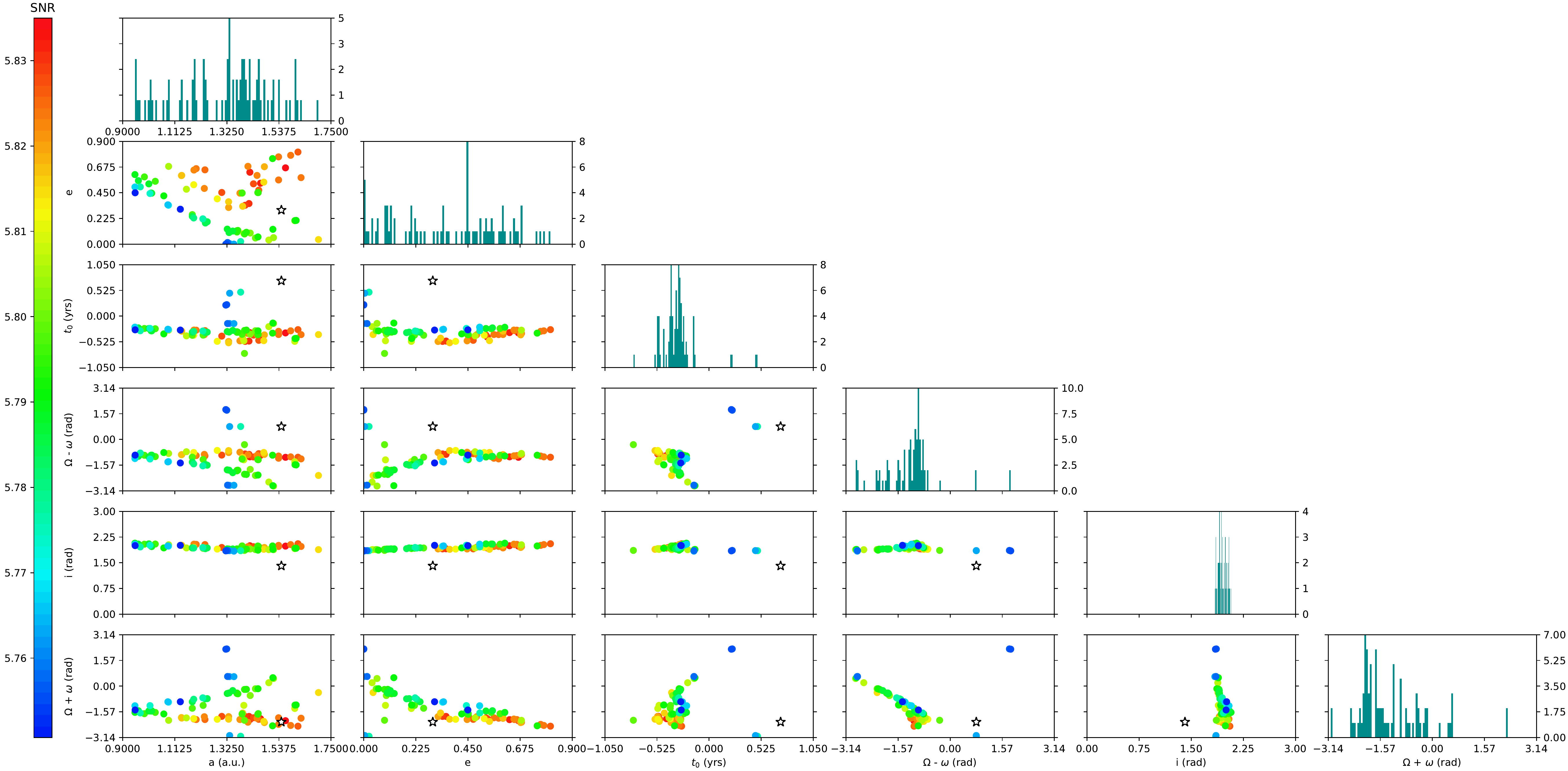}
\caption{Same as Fig~\ref{fig:coner_plot_0.5Yr_orbit15} but using the true epochs of observations as reported in \citetads{2021NatCo..12..922W}}
\label{fig:coner_plot_epochs_wagner_orbit15}
\end{figure*}

When the observations span a significant part of the orbit (6 months), the histograms of the orbital parameters have their peaks well centered on the injected values (see histograms and star symbol in Fig. \ref{fig:coner_plot_0.5Yr_orbit15}). Howerver, when the observations only span a small fraction of the orbit (1 month), the correct orbital parameters are not well recovered (except for $t_0$, see the star symbol that is not at the position of the histogram peaks in Fig. \ref{fig:coner_plot_epochs_wagner_orbit15}).

\end{appendix}

\end{document}